\documentclass[prd]{revtex4}

\usepackage{amsbsy}
\usepackage{amssymb}
\usepackage{amsmath}
\usepackage{natbib}
\usepackage{rotating}
\usepackage{times}
\usepackage{multirow}

    \newcommand{\ncd}{\newcommand}
    \ncd{\mrm}    {\mathrm}
    \ncd{\beq} {\begin{equation}}
    \ncd{\eeq} {\end{equation}}

\def\n{{\rm f}}

    \def\n{{\rm n}}
    \def\s{{\rm s}}

    \ncd{\nns}{n_{\n\s}^2}
    \ncd{\wns}{w_a^{\n\s}}
    \ncd{\us}{u_a^\s}
    \ncd{\starq}{\star{\bf q}}

    \ncd{\thetastar}{\theta^*}
    \ncd{\sstar}{s^*}

    \ncd{\un}{u_a^\n}
    \ncd{\Bn}{\mathcal{B}^\n}
    \ncd{\Bs}{\mathcal{B}^\s}
    \ncd{\Ans}{\mathcal{A}^{\n\s}}
    \ncd{\Ann}{\mathcal{A}^{\n\n}}
    \ncd{\Ass}{\mathcal{A}^{\s\s}}
    \ncd{\Xns}{\chi^{\n\s}}
    \ncd{\Asn}{\mathcal{A}^{\s\n}}
    \ncd{\Xsn}{\chi^{\s\n}}
    \ncd{\nn}{\nonumber}

\ncd{\dell}{\partial}
\ncd{\nfrac}[2]{\left(\frac{n_{#1}}{n_{#2}}\right)^2}
\ncd{\pc}{\check{p}}
\ncd{\rhoc}{\check\rho}
\ncd{\betac}{\check\beta}
\ncd{\muc}{\check\mu}
\ncd{\Oc}{\check\Omega}
\ncd{\ec}{\check\epsilon}
\ncd{\tsfrac}[2]{{\textstyle\frac{#1}{#2}}}

\begin{document}

\title{A consistent first-order model for relativistic heat flow}

\author{Nils Andersson \& Cesar Lopez-Monsalvo}

\affiliation{Mathematical Sciences, University of Southampton, UK}

\begin{abstract}
This paper revisits the problem of heat conduction in relativistic fluids, associated with issues concerning both stability and causality.
It has long been known that the problem requires information involving second order deviations from thermal equilibrium. 
Basically, any consistent first-order theory needs to remain cognizant of its higher-order origins. We demonstrate this by carrying out the required first-order reduction of a recent variational model. We provide an analysis of the dynamics of the system, obtaining the conditions that must be satisfied in order to avoid instabilities and acausal signal propagation. The results demonstrate, beyond any reasonable doubt,  that the model has all the features one would expect of a real physical system. In particular, we highlight the presence of a second sound for heat in the appropriate limit. We also make contact with previous work on the problem by showing how the various constraints on our system agree with previously established results. 
\end{abstract}

\maketitle

\section{Context}

Relativistic thermodynamics continues to provide interesting challenges, in particular in the context of dissipative and nonlinear
phenomena. The issues involved range from direct applications in various areas of physics to fundamental problems like the nature of time (visavi the second law of thermodynamics) and the formation of structures at nonlinear deviations from thermal equilibrium.
Much recent work has been motivated by the modelling of complex astrophysical systems, like neutron stars \cite{livrev},
and cosmology \cite{maartens}. There has also been a resurgence of interest in dissipative systems
in the context of colliders like RHIC at Brookhaven and the LHC at CERN \cite{rhic1,rhic2,rhic3}.
These latter developments, which have to a large extent been driven by the need to understand the dynamics of a hot quark-gluon plasma, are often linked with underlying principles like the AdS/CFT conjecture and holography \cite{rangamani}. Even though the problem dates back to the origins of relativity theory, it remains (in a slightly different guise) at the forefront of modern thinking. 

According to the established consensus view, 
one must account for second-order deviations from thermal equilibrium in order to achieve causality and stability.  This is 
certainly the lesson from the celebrated work of Israel and Stewart \cite{IS,IS2}, see \cite{isrec1,isrec2,isrec3} for recent work on the problem. We have recently revisited the key points in the context of heat conduction \cite{cesar}, taking a multi-fluid prescription based on  Carter's convective variational formulation for relativistic fluids \cite{carter02}
as our starting point. This is a mathematically elegant approach that has the flexibility required to account for the 
physics that we need to consider. A particularly appealing feature of the variational approach is that, once an ``equation of state'' for matter is provided, the theory provides
the relation between the various currents and their conjugate momenta \cite{livrev}. The variational analysis leads to a second-order model which has the key elements required for causality and stability, in particular, it clarifies the role of the inertia of heat (e.g., the effective mass associated with phonons). This effect enters the model in an intuitive fashion in terms of entrainment between the matter and heat \cite{nilsclass}. As demonstrated by Priou \cite{priou} some time ago, the final variational model is formally equivalent to the Israel-Stewart construction. 
This exercise demonstrates clearly that the relaxation associated with causal heat transport is determined by the thermal inertia. At the end of the day, the theoretical framework becomes rather intuitive and the physics involved seems natural.

Does this mean that no troublesome issues remain in this problem area? Not quite. First of all, it is clear that the need to introduce additional parameters (e.g., the relevant relaxation times) and keep track of higher order terms (fluxes of the fluxes etcetera) make actual applications rather complex. Secondly, we are not much closer to considering systems that deviate significantly from equilibrium, such that there is no natural ``small'' parameter to expand in. The variational model sheds some light on this regime by clarifying the role of the temperature in systems out of equilibrium, but there is some way to go before we understand issues associated with, for example, any ``principle of extremal entropy production'' and instabilities that lead to structure formation. Finally, despite the obvious successes of the extended thermodynamics framework \cite{joubook}, there is no universal agreement concerning the validity (and usefulness) of the results. To some extent this is natural given the interdisciplinary nature of the problem; to make progress we need to account for both 
thermodynamical principles and fundamental general relativity. This leads to a range of deep questions concerning, in 
particular, the actual meaning of the variables involved in the different models (e.g., the entropy). The ultimate theory must have a clear link with statistical physics and even information theory. Our efforts are not yet at that level. Basically, we need to continue to make progress if we are to address fundamental problems in, for example, cosmology.

This paper sets a rather more modest target; we want to explore the extent to which a ``first-order'' formulation for heat conduction in general relativity is viable. The question may seem somewhat odd given that we have already acknowledged the need to account for (at least) second order contributions. However, it is interesting to ask whether a first-order model
may nevertheless be useful (possibly in a somewhat restricted sense). We will demonstrate that this is, indeed, the case.
Noting that the original first order models, due to Eckart \cite{eckart} and Landau and Lifshitz \cite{landau}, were incomplete we develop a
consistent framework that includes the key thermal relaxation. We then consider the properties of this model, and 
show that it can be made both stable and causal (making contact with the classic work by Hiscock and Lindblom \cite{hl1,hl2}). This does not mean that the system may not exhibit instabilities. On the contrary, instabilities are in a sense generic in these problems \cite{2stream}, but the analysis sheds further light on the nature of these instabilities and also elucidates the stabilizing role of the thermal inertia. The discussion also provides insight into the emergence of second sound (an effect that has been experimentally
verified in low temperature crystals) associated with the heat transport. This provides a key link to systems that exhibit superfluidity, and demonstrates the potential for a unified treatment of heat transport in normal and superfluid matter. 

\section{Thermal dynamics}

We take as our starting point our recent variational analysis of the relativistic heat problem \cite{cesar}.
The model is phenomenological, and assumes that the entropy component can be treated as a ``fluid". In essence, 
this implies that  the mean free path of the phonons 
is taken to be small compared to the model scale. We then consider two fluxes, one corresponding to the matter flow
and one which is associated with the entropy. The latter is essentially treated as a massless (zero rest-mass) fluid. The 
dynamics then follows from a Lagrangian which depends on the relative flow of the two fluxes.
The associated entrainment
turns out to be a crucial feature of the problem \cite{nilsclass,carter03,cesar}.

We
assume that the particle number is large enough that the fluid approximation applies and  there is a well defined matter current, $n^a$.
Moreover, we adopt the multi-fluid view and treat the entropy as an effective fluid with flux $s^a$. This current is in general not aligned with the particle flux.
The misalignment is associated with the heat flux and leads to entropy production.

As in the case of a generic two-fluid system (see \cite{carterlanglois} for an example in the case of a cool relativistic superfluid), the starting point is the definition of a relativistic invariant Lagrangian $\Lambda$.
Assuming that the system is isotropic, we take $\Lambda$ to be a function of the different scalars that can be formed
by the two fluxes~\footnote{The natural way to account for viscosity is to allow
the master function to depend also on the associated stresses \cite{cadis}.}. From $n^a$ and $s^a$ we can form three scalars;
    \beq
    n^2  = -n_a n^a \ , \quad
    s^2  = -s_a s^a \ , \quad
    j^2  = -n_a s^a \ .
    \eeq
An unconstrained variation of $\Lambda$ then leads to
    \beq
    \label{var1}
    \delta \Lambda = \frac{\partial \Lambda}{\partial n}\delta n + \frac{\partial \Lambda}{\partial s} \delta s + \frac{\partial \Lambda}{\partial j} \delta j \ .
    \eeq
Changing the passive density variations for dynamical variations of the worldlines generated by the fluxes and the metric (as discussed in \cite{livrev}) we find that
  \beq
    \label{var2}
    \delta \Lambda = \left[-2\frac{\partial \Lambda}{\partial n^2}n_a -\frac{\partial \Lambda}{\partial j^2}s_a \right]\delta n^a+
                \left[ -2\frac{\partial \Lambda}{\partial s^2}s_a -\frac{\partial \Lambda}{\partial j^2}n_a \right]\delta s^a  
           + \left[-\frac{\partial \Lambda}{\partial n^2}n^an^b - \frac{\partial \Lambda}{\partial s^2}s^a s^b - \frac{\partial \Lambda}{\partial j^2}n^a s^b\right]\delta g_{ab} \ .
           \eeq
From this result we can read off the conjugate momentum associated with each of the fluxes;
    \beq
    \mu_a=\frac{\partial \Lambda}{\partial n^a} =  g_{ab}(\Bn n^b + \Ans s^b) \ , \quad
    \theta_a=\frac{\partial \Lambda}{\partial s^a} =  g_{ab}(\Bs s^b + \Ans n^b) \ ,
    \eeq
where we have introduced the coefficients \cite{livrev,cesar}
    \beq
    \label{var.coefs}
    \mathcal{B}^\n\equiv -2 \frac{\partial \Lambda}{\partial n^2}, \quad \mathcal{B}^\s\equiv -2 \frac{\partial \Lambda}{\partial s^2}, \quad \mathcal{A}^{\n\s}\equiv-\frac{\partial \Lambda}{\partial j^2} \ .
    \eeq

The energy-momentum tensor is obtained by noting that the displacements of the conserved currents induce
a variation in the spacetime metric and therefore the variations of the fluxes, $\delta n^a$ and $\delta s^a$,
are constrained.  The energy-momentum tensor is thus found to be
    \beq
    \label{se-tensor}
    T_a^{\ b} = \mu_a n^b + \theta_a s^b + \Psi \delta_a^{\ b} \ ,
    \eeq
where we have defined the generalized pressure, $\Psi$, as
    \beq
    \label{psi}
    \Psi = \Lambda -\mu_a n^a - \theta_a s^a \ .
    \eeq

As a result of the coordinate invariance associated with general relativity, the divergence of the energy-momentum tensor (\ref{se-tensor}) vanishes.
For an isolated system, we can express this requirement as an equation
of  force balance
    \beq
    \label{fbal}
    \nabla_b T_{a}^{\ b} = f^\n_a + f^\s_a = 0 \ ,
    \eeq
where the individual force densities are \cite{cesar}
    \begin{align}
    \label{fn}
    f^\n_a &=2 n^b\nabla_{[b}\mu_{a]} + \mu_a \nabla_b n^b \ , \\
    \label{fs}
    f^\s_a &=2 s^b  \nabla_{[b}\theta_{a]}+ \theta_a \nabla_b s^b \ .
    \end{align}

We note that, in order to obtain the energy momentum tensor (\ref{se-tensor}) we needed to impose the conservation of the fluxes as constraints on the variation \cite{livrev}.
However, the equations of motion, (\ref{fn}) and (\ref{fs}), still allow for non-vanishing  production terms.
If we, for simplicity, consider a single particle species, the matter current is conserved and we have
$\nabla_a n^a = 0$.
This removes the second term from the right-hand side of (\ref{fn}). In contrast, the entropy flux is  generally not conserved. In accordance with the second law, we must have
\beq
\label{divs}
\nabla_a s^a = \Gamma_\s \ge 0 \ .
\eeq

\subsection{Temperature}

To make progress, we need to connect the general variational results with the relevant
thermodynamical concepts. In doing this it makes sense to consider a specific choice of frame.
In the context of a single (conserved) species of matter, we see that force $f^\n_a$ is orthogonal to the matter flux, $n^a$, and therefore it has only three degrees of freedom. Furthermore, because of the
force balance \eqref{fbal}, we also have $n^a f^\s_a=0$. This suggests that it is natural to focus on observers associated with the matter frame. We therefore introduce the four-velocity $u^a$ such that $n^a = n u^a$,
where $u_a u^a = -1$ and $n$ is the number density measured in this frame.

Having chosen to work in the matter frame (in the spirit of Eckart \cite{eckart}), we can decompose the entropy current and its conjugate momentum into parallel  and orthogonal components. The entropy flux is then expressed as
    \beq
    s^a =  s^* (u^a + w^a) \ ,
    \eeq
where $w^a$ is the relative velocity between the two fluid frames, and $u^a w_a=0$. Letting $s^a = s u^a_\s$ where $u_\s^a$ is the four-velocity associated with the
entropy flux, we see that $s^*=s\gamma$ where $\gamma$
is the redshift associated with the relative motion of the two frames~\footnote{In the following, we will use an asterisk to denote matter frame quantities.}. This illustrates the subjective nature of entropy. It is an observer dependent quantity, not an absolute notion.

Similarly, we can write the thermal momentum as
    \beq
    \theta_a = \thetastar u_a + \Bs s^*  w_a = \left(\Bs s^* + \Ans n\right) u_a + \Bs s^* w_a \ .
    \eeq
This leads to a measure of the temperature
measured in the matter frame;
\beq
-u^a \theta_a = \thetastar =  \Bs s^* + \Ans n \ .
\label{tpar}\eeq
In essence, this quantity represents the effective mass of the entropy component.
Returning to the stress-energy tensor, and making use of the projection orthogonal to the matter flux, we find that
the heat flux (energy flow relative to the matter) is given by
    \beq
    \label{heat}
    q_a = -\perp_{ab}u_c T^{bc} = s^* \thetastar w_a \  , 
    \eeq
    where we have used the projection
 \beq
 \perp^{ab} = g^{ab} + u^a u^b \ .
 \eeq
Defining the new  variables $\sigma^a = s^* w^a$ and $p_a = \Bs s^* w_a$, the energy density  measured in the matter frame can be obtained by a Legendre-type transform
on the master function. That is, we have
    \beq
    \label{rhostar}
    \rho^* = u_a u_b T^{ab} = - \Lambda + p_a \sigma^a \ .
    \eeq

The relevance of the new variables becomes apparent if we consider the fact that  the \emph{dynamical} temperature
in \eqref{tpar}  agrees with the  \emph{thermodynamical} temperature that an observer moving with the matter would measure. In other words, we have
\beq
\theta^* = \left. {\partial \rho^* \over \partial s^*}\right|_{n,p} \ ,
\eeq
where $\rho^* = \rho^*(n,s^*,p)$. This is, of course,
the standard definition of temperature as energy per degree of freedom of the system.
Mathematically, the temperature is obtained from the variation of the energy with respect to the entropy in the observer's frame (keeping the other thermodynamic variables fixed).

This result is not trivial. The requirement that the two temperature measures agree
determines the additional state parameter, $p$, to be held constant in the variation of $\rho^*$.
The importance of the chosen state variables is emphasized further if we note that, when the system is out of equilibrium, the energy depends on the heat flux
(encoded in $\sigma^a$ and $p_a$).
This leads to an \emph{extended} Gibbs relation (similar to that postulated in many approaches to extended thermodynamics \cite{joubook});
  \beq
    d \rho^* = \mu d n + \theta^* d s^* + \sigma d p \ .
    \eeq
This result arises naturally from the variational analysis.

According to the traditional view, thermodynamic properties like pressure and temperature are uniquely defined only in equilibrium.
Intuitively this makes sense since, in order to carry out a measurement, the measuring device must have time to
reach ``equilibrium'' with the system. A measurement is  only meaningful as long as the timescale required to obtain a result is
shorter than the evolution time for the system. However, this does not prevent a generalisation of the various
thermodynamic concepts (as described above). The procedure may not be ``unique'', but one should at least require the generalised concepts to be internally consistent
within the chosen extended thermodynamics model. Our model satisfies this criterion.

\subsection{Causal heat flow}

The variational model encodes the finite propagation speed for heat, as required by causality.
To see this, we use
the orthogonality of the entropy force density $f_\s^a$ with the
matter flux,  solve for the entropy production rate $\Gamma_\s$  and finally impose the second law of thermodynamics.
It is natural to express the result in terms of the heat flux $q^a$, defined by
    \beq
    \label{sdecomp}
    s^a= s^*u^a + \frac{1}{\thetastar}q^a \ .
    \eeq
We also let the conjugate momentum takes the form
    \beq
    \label{thetadec}
    \theta_a = \thetastar u_a + \beta q_a \ ,
    \eeq
where we have defined
    \beq
    \label{beta}
    \beta = \left(\frac{1}{s^*} -  \frac{\Ans n }{s^* \thetastar} \right) \ .
    \eeq
With these definitions, we  impose the second law of thermodynamics by demanding that the entropy production is a quadratic in the sources, i.e.,
\beq
\Gamma_\s = {q^2 \over \kappa \theta_*^2}  \ge 0 \ ,
\label{gs1}\eeq
where   $\kappa>0$ is the the thermal conductivity.
This means that the heat flux is governed by
    \beq
    \label{gato}
    \tau \left( \dot{q}^a + q_c \nabla^a u^c \right) + q^a = -\tilde\kappa \perp^{ab}\left( \nabla_b \thetastar + \thetastar \dot u_b\right) \ ,
    \eeq
    where $\dot{q}^a = u^b \nabla_b q^a$ and $\dot{u}^a$ is the four-acceleration (in the following, dots represent time derivatives in the matter frame).
  Here we have introduced
    \beq
    \tilde\kappa \equiv \frac{\kappa}{1 + \kappa \dot \beta} \ ,
    \eeq
while the thermal relaxation time is given by
    \beq
    \tau = \frac{\kappa\beta}{1 + \kappa \dot \beta} \ .
    \eeq
The final result (\ref{gato}) is the relativistic version of the so-called Cattaneo equation \cite{cesar}.
From the analysis we learn that the entropy entrainment, encoded in $\Ans$, plays a key role in determining the thermal relaxation time $\tau$.
This agrees with the implications of extended thermodynamics, and echoes recent results in the context of Newtonian gravity \cite{nilsclass}.

\subsection{The matter flow}

The heat problem has two dynamical degrees of freedom. So far, we have focussed on the entropy. In addition to the relativistic Cattaneo equation
\eqref{gato} we have
a momentum equation for the matter component. From \eqref{fn} it follows that this equation can be written
\beq
 \mu \dot{u}_a+\perp^b_a\nabla_b \mu +\alpha \dot{q}_a +\dot{\alpha} q_a +\alpha q^b \nabla_a u_b = {1 \over n} f^\n_a \ .
\label{mom}\eeq
Here we have represented the matter momentum by
\beq
\mu_a = \mu u_a + \alpha q_a \ ,
\eeq
where $\mu$ is the chemical potential (in the matter frame) and 
\beq
\alpha = {\Ans \over \theta^*} \ .
\label{alp}\eeq
This means that we have
\beq
\alpha = {1 - \beta s^* \over n} \ .
\eeq
Given these definitions, we have [c.f., \eqref{gs1}]
\beq
-f^\n_a = f^\s_a = - {1 \over \tilde{\kappa}}\left[ s^* - {\beta q^2 \over \left(\theta^*\right)^2} \right] q_a  \ .
\eeq
It is useful to note that this implies that the force has a term that is linear in $q^a$. This will be important later.

These relations complete our summary of the heat conduction model developed in \cite{cesar}.

\section{A consistent first-order model}

The model we have described crucially contains terms that enter at second order of deviation from thermal equilibrium, e.g., terms that are second order in the heat flux $q^a$. Moreover, it is clear that key effects (like the entropy entrainment) arise from the presence of 
second order terms in the Lagrangian $\Lambda$. Having said that, it is obviously the case that we can truncate the model at first order.
This does \underline{not} take us back to the original first-order model discussed by Eckart \cite{eckart}. Crucially, the thermal relaxation 
remains. Basically, this reflects the simple fact that you need to know the energy of a system to quadratic order in order to develop the complete linear equations of motion. Noting this, it is interesting to consider the features of this new first-order model. First of all, we can expect to get a clearer understanding of some of the general features of the variational model. Secondly, we may also
find that this, much simpler, model is adequate for many situations of practical interest.

\subsection{The linear model}

We want to restrict our analysis to first order deviations from equilibrium.  
Thermal equilibrium corresponds to $q^a=0$, no heat flux, and $\dot{u}^a=0$, no matter acceleration. 
Moreover, in the simplest cases there should be no shear, divergence or vorticity associated with the flow, i.e., we will have $\nabla_a u^a = 0$ and 
$\nabla_b u^a=0$ as well~\footnote{There are obviously many relevant problems that require a more 
``complicated'' equilibrium, e.g. rotating stars and an expanding universe in cosmology. It is, however, straightforward to extend our analysis to these cases.}. Treating all these quantities as being of first order, and noting that 
\beq
u_b \dot{q}^b = - q^b \dot{u}_b \ ,
\eeq
contributes at second order, we arrive at  two momentum equations; from \eqref{mom} we have
\beq
\mu \dot{u}_a + \perp^b_{\ a} \nabla_b \mu + \alpha \dot{q}_a + \left( \dot{\alpha} - {s \over n\tilde{\kappa}} \right) q_a  =  0 \ ,
\label{umom}\eeq
while \eqref{gato} leads to 
\beq
\tau \dot{q}_a + q_a + \tilde{\kappa} \left(  \perp^b_{\ a} \nabla_b T + T \dot{u}_a \right)= 0  \ ,
\label{qmom}\eeq
We also have the 
two conservation laws
\beq
\nabla_a n^a = 0 \ ,
\eeq
\beq
\nabla_a s^a = 0 \ .
\eeq
In these equations we have used the fact that $s^*$ and $\theta^*$ differ from the equilibrium values $s$ and $T$ only at second order.
Moreover, to first order the pressure $p$ is obtained from the standard equilibrium Gibbs relation;
\beq
\nabla_a p = n \nabla_a \mu + s \nabla_a T \ .
\label{pres}\eeq
Finally, we have the fundamental relation
\beq
\rho + p = \mu n + s T \ .
\label{fund1}\eeq
By comparing \eqref{umom} and \eqref{qmom} to Eckart's results it becomes apparent 
 to what extent the first-order model remains cognizant of its higher order origins.
Specifically, $\alpha$ and (therefore) $\tau$ depend on $\Ans$ and the entropy entrainment, c.f., \eqref{alp}. These effects rely on quadratic terms in the Lagrangian, and hence would not be present in a model that includes only first order terms from the outset. Hence, they are absent in Eckart's model.

In order to analyze the dynamics of the heat problem, we will consider perturbations (represented by $\delta$) away from a uniform
equilibrium state. First of all, we have  $q_a=\dot{u}_a=0$ for a system in equilibrium. We can also ignore $\dot{\alpha}$ and $\dot{\beta}$, since the equilibrium configuration is uniform, which means that we can replace $\tilde{\kappa}$ by $\kappa$. 
This means that we are left with the two equations;
\beq
\mu \delta \dot{u}_a + \perp^b_{\ a} \nabla_b \delta \mu + \alpha \delta \dot{q}_a-{s \over n\kappa} \delta q_a= 0 \ ,
\label{eq1}\eeq
and
\beq
\tau \delta \dot{q}_a + \delta q_a + \kappa \perp^b_{\ a} \nabla_b \delta T + \kappa T \delta \dot{u}_a = 0  \ ,
\label{eq2}\eeq

It is worth noting that we can combine these two to get
\beq
\left( p+\rho\right)\delta \dot{u}_a + \perp_a^b \nabla_b \delta p + \delta\dot{q}_a = 0  \ .
\label{eq3}\eeq
The last two equations [\eqref{eq2} and \eqref{eq3}] are, not surprisingly, identical to the first-order reduction of the Israel-Stewart model. 
This means that the problem is relatively well explored. In particular, the conditions required for stability and causality were 
derived by  Hiscock and Lindblom \cite{hl1,hl2}, see also Olson and Hiscock \cite{olson}, quite some time ago. 
However, there are good reasons to revisit the problem. Most importantly, 
there is clear evidence from the recent literature (c.f., discussions of the relevance of the thermal relaxation and the role of the coupling between the four acceleration and the heat flux \cite{heatcoup1,heatcoup2,heatcoup3}) that the key lessons from almost three decades ago have not been appreciated. To some extent this could be due to the fact that the 
Hiscock-Lindblom analysis is rather involved. Our aim is to clarify the main issues in the simpler context of heat conduction (ignoring 
viscosity). We also want to emphasize aspects that were only mentioned in passing in early work. Particularly relevant in this respect is the 
existence of second sound; an effect that is prominent in superfluids but which has also been observed in low-temperature crystals.
We will demonstrate how the second sound emerges within the causal heat-conduction model. The overarching aim is to establish, beyond any reasonable doubt, that the model represented by \eqref{eq2} and either \eqref{eq1} or \eqref{eq3} has all the properties expected of a 
reliable model for heat conduction in general relativity.

\subsection{Transverse waves: Stability}

Working in the frame associated with the background flow, we note that \eqref{eq1} and
\eqref{eq2} only have spatial components. That is, we may erect a local Cartesian coordinate system associated with the matter frame and 
simply replace $a\rightarrow i$ where $i=1-3$.
Then taking the curl ($\epsilon^{jki} \nabla_k$) of the equations in the usual way, we arrive at
\beq
m_\star \dot{U}^i - {1 \over \tau}  \dot{Q}^i = 0 \ ,
\eeq
and
\beq
m_\star \dot{Q}^i + (p+\rho) Q^i  = 0 \ ,
\eeq
where we have defined
\beq
U^i = \epsilon^{ijk} \nabla_j \delta u_k \ , \qquad \mbox{and} \qquad Q^i = \epsilon^{ijk} \nabla_j \delta q_k \  , 
\eeq
and
\beq
m_\star = n \left( \mu - { \alpha \kappa T \over \tau} \right) = p+ \rho - {\kappa T \over \tau} \ .
\eeq

Assuming that the perturbations depend on time as $e^{i\omega t}$, where $t$ is the time-coordinate associated with the matter frame,
 we arrive at the dispersion relation
for transverse perturbations;
\beq
i\omega \left[  (p+\rho)( 1 + i\omega \tau) - i \omega \kappa T \right]= 0 \ .
\eeq
Obviously $\omega = 0$ is a solution. The second root is
\beq
\omega =  {i(p+\rho) \over m_*\tau} \ .
\eeq
This results shows that the thermal relaxation time $\tau$ is essential in order for the system to be stable.
We need $m_*>0$, i.e., the relaxation time must be such that
\beq
\tau > {\kappa T \over p+\rho}  \ .
\label{tcon}\eeq
The analysis clearly shows that Eckart's model (for which $\tau=0$) is inherently unstable. 
Moreover, the constraint on the relaxation time agrees with one of the conditions obtained by Olson and Hiscock \cite{olson} (c.f., their eq. (41)), representing the inviscid limit of the exhaustive analysis of the Israel-Stewart model of Hiscock and Lindblom \cite{hl1}. We may also note that the condition given in 
eq. (43) of \cite{olson} simply leads to the weaker requirement $\tau \ge 0$.

The physical origin of the transverse instability can be understood if  rewrite \eqref{eq2} and \eqref{eq3} as
\beq
m_\star \delta \dot{u}_a + \perp^b_a \nabla_b \delta p - {1 \over  \tau} \left( \delta q_a + \kappa \perp^b_a\nabla_b  \delta T \right)
= 0 \ ,
\eeq
and
\beq
m_\star \delta \dot{q}_a + (p+\rho) \left( \delta q_a + \kappa \perp^b_a \delta T \right) -  \kappa T \perp^b_a \nabla_b \delta p = 0  \ .
\eeq
These relations show that $m_*$ plays the role of an ``effective'' inertial mass (density).
The importance of this quantity has been discussed in a series of papers by Herrera and collaborators \cite{herr1,herr2,herr4}, especially in the context of gravitational collapse. Basically, the  instability of the Eckart formulation is due to the inertial mass of the fluid becoming negative. 
Once this happens the pressure gradient no longer provide a restoring force, rather it would tend to push the system further away from equilibrium. 
This is a run-away process, associated with exponential growth of the perturbations. Ultimately, the instability is due to the inertia of heat; an unavoidable consequence of the equivalence principle (heat carries energy, which means that it can be associated with an effective mass \cite{tolman}). The condition \eqref{tcon} may seem rather extreme (Hiscock and Lindblom \cite{hl2} quote a timescale of $10^{-35}$~s for water at 300K), but it sets a sharp lower limit for the thermal relaxation in physical systems. A system with faster thermal relaxation can not settle down to equilibrium. However, it may still be reasonable to ask if a system may evolve in such a way that it enters the 
unstable regime (in the way discussed in \cite{herr1,herr2}). Given our assumed equilibrium the present formulation does not allow us to consider this 
question, but it seems clear that if a system were to evolve in that way then one would need a fully nonlinear analysis to determine the 
consequences. 

\subsection{The longitudinal problem}

The transverse problem is relatively simple since the are no corresponding restoring forces in  a simple fluid problem (these requires rotation, elasticity, the 
presence of a magnetic field etcetera). When we turn to the longitudinal case the situation changes. In a perfect fluid longitudinal perturbations propagate as sound waves, and when we add complexity to the model the dispersion relation can get very complicated. 

To analyze the longitudinal case, we take the divergence of \eqref{eq2} and 
\eqref{eq3}. This leads to
\beq
(p+\rho) \partial_t \nabla_i \delta u^i + \nabla^2 \delta p +  \partial_t \nabla_i \delta q^i = 0 \ , 
\eeq
and
\beq
\tau \partial_t \nabla_i \delta q^i + \nabla_i \delta q^i + \kappa \nabla^2 \delta T + \kappa T \partial_t \nabla_i \delta u^i = 0 \ .
\eeq
We also need the conservation laws which take the form
\beq
\nabla_i \delta u^i = - {1 \over n} \partial_t \delta n \ , 
\eeq
\beq
\nabla_i \delta q^i = - nT \partial_t \delta \bar{s} \ ,
\eeq
where we have defined the specific entropy; $\bar{s} = s/n$. To make progress we need to, first of all, decide what variables to work with, and secondly we need 
an equation of state for matter. In the following we will opt to work with the perturbed densities $\delta n$ 
and $\delta \bar{s}$. Keeping in mind that we are only retaining first-order quantities, we have
\beq
\delta p = \left( {\partial p \over \partial n } \right)_{\bar{s}} \delta n + \left( {\partial p \over \partial \bar{s}} \right)_n \delta \bar{s} \ ,
\eeq
and
\beq
\delta T = \left( {\partial T \over \partial n} \right)_{\bar{s}} \delta n + \left({\partial T \over \partial \bar{s}} \right)_n \delta \bar{s} \ .
\eeq
It is also useful to keep in mind that the temperature represents the ``entropy chemical potential'', i.e. is 
defined by  
\beq
T = \left({\partial \rho \over \partial s} \right)_n \ ,
\eeq
where $\rho=\rho(n,s)$ represents the equation of state. This immediately implies that 
\beq
\left({\partial T \over \partial n} \right)_s = \left({\partial \mu \over \partial s} \right)_n \ , 
\eeq
i.e., we can  reduce the number of thermodynamic quantities.  In order to make contact with (potential) observations, it is natural to work with (i) 
the adiabatic speed of sound;
\beq
c_s^2 = \left({\partial p \over \partial \rho} \right)_{\bar{s}} = {n\over p+\rho}  \left( {\partial p \over \partial n } \right)_{\bar{s}}  \ ,
\eeq
(ii) the heat capacity at fixed volume;
\beq
c_v = T \left( {\partial \bar{s} \over \partial T} \right)_{n} \ ,
\eeq
and (iii)
\beq
\alpha_s = {n\over T} \left( {\partial T \over \partial n} \right)_{\bar{s}} = {T\over n} \left( {\partial p \over \partial \bar{s}} \right)_{n} \ .
\eeq
For future reference, it is also useful to note the identity [c.f., eq. (96) in Hiscock and Lindblom \cite{hl1}]
\beq
{1\over c_v} - {1\over c_p} = {n^3 \over T (p+\rho) c_s^2 }  \left( {\partial T \over \partial n} \right)_{\bar{s}}^2 = {nT \over (p+\rho) c_s^2} \alpha_s^2 \  ,
\label{cvrel}\eeq
where $c_p$ is the heat capacity at fixed pressure.

If we (again) focus on plane-wave solutions such that the perturbations behave as $\exp(i\omega t + i k_j x^j)$,
and introduce the phase-velocity $\sigma = \omega/k$, then the above relations lead to the coupled equations
\beq
{p+\rho \over n} \left( \sigma^2 -c_s^2\right) \delta n + n T \left( \sigma^2 - 
\alpha_s \right) \delta \bar{s} = 0 \ , 
\eeq
and
\beq
{\kappa T\over n} \left( \sigma^2 -\alpha_s\right) \delta n  
 + T \left( n \tau  \sigma^2 - {i \sigma n \over k} - {\kappa \over c_v} \right) \delta \bar{s} = 0 \ .
\eeq
From these results it is easy to see that the generic dispersion relation will be a quartic in $\sigma$. This 
is exactly what one would expect for a physical system. As we have already mentioned, the formulation must accommodate the presence
of ``second sound'' which has been experimentally verified for both superfluids and low temperature solids. 
Working out the explicit dispersion relation, we find
\beq
m_*\tau \sigma^4 - {i(p+\rho) \over k} \sigma (\sigma^2 - c_s^2)  
- \left[ (p+\rho)\left( {\kappa\over n c_v} + c_s^2 \tau \right)-2\kappa T \alpha_s\right] \sigma^2  +  \kappa \left[ {p+\rho\over n} {c_s^2 \over c_{v}}  - T\alpha_s^2\right] = 0 \ .
\label{quartz}
\eeq
This expression is still too complicated for us to be able to make definite statements about the solutions, 
without making further assumptions. The most direct strategy would be to consider an explicit equation of state, work out the relevant thermodynamics quantities, and then solve \eqref{quartz} (probably numerically). This would allow us to establish whether the considered model is stable and causal. This route is, however, not particularly 
attractive given the need to introduce an explicit model. If we want to continue to consider the problem 
in (at least to some extent) generality, then we need to resort to approximations. As we will see, this is 
a very instructive route.

In order to simplify the analysis, we will consider the long- and short-wavelength limits of the problem. The results we obtain in these limits provide 
useful illustrations of the key features. At the same time, we should keep in mind that
both cases are somewhat ``artificial''. First of all, fluid dynamics is, fundamentally, an effective long-wavelength theory in the 
sense that it arises from an averaging over a large number of individual particles (constituting each fluid element).
In effect, the model only applies to phenomena on scales much larger than (say) the interparticle 
distance. However, the infinite wavelength limit represents a uniform system, which is  artificial since real physical systems tend to be finite.
Moreover, as we will not  account explicitly for gravity  we can only consider scales 
on which spacetime can be considered flat. While the plane-wave analysis holds on arbitrary scales in special relativity, a curved spacetime 
introduces a cut-off lengthscale beyond which the analysis is not valid. The theoretical framework is valid in general, but 
on larger scales we would have to consider also the perturbed Einstein equations. 

\subsection{The long wavelength problem}

Let us first consider the long wavelength, $k\to0$, problem. This represents the true hydrodynamic limit, and it   easy to see that there are two sound-wave solutions and two modes that are predominantly
diffusive. The sound-wave solutions take the form
\beq
\sigma \approx \pm c_s \left[1 \pm i {\kappa T \over 2(p+\rho) c_s^3} (c_s^2 - \alpha_s)^2 k \right]  \ .
\eeq
These solutions are clearly stable, since Im~$\sigma>0$. Using the Maxwell relations listed by Hiscock and Lindblom \cite{hl1}, we can show
that this results agrees with eq. (40) from \cite{hl2}. Moreover, our result simplifies to [using \eqref{cvrel}]
\beq
\mathrm{Im}~\sigma \approx {\kappa \over 2n} \left( {1 \over c_v} - {1 \over c_p} \right) \ ,
\label{nonrel}\eeq
in the limit where $|\alpha_s|\gg c_s^2$, which is relevant since $c_s^2 \sim p/\rho$ becomes small in the non-relativistic limit. Indeed, we find that 
\eqref{nonrel}
agrees with the standard result for sound absorption in a heat-conducting medium \cite{mountain}.

In addition to the sound waves, we have a slowly damped solution
\beq
\sigma \approx i\kappa \left[ {1 \over n c_v} - {T \alpha_s^2 \over (p+\rho) c_s^2 } \right] = {i\kappa \over n c_p}  \ .
\eeq
This is the classic result for thermal diffusion.

Finally, the system has a fast decaying solution;
\beq
\sigma \approx  {i (p+\rho) \over m_*  k \tau}  \ .
\eeq
Under most circumstances, this root decays too fast to be observable. This means that the model reproduces that standard 
``Rayleigh-Brillouin spectrum'' 
with two sound peaks symmetrically placed with respect to the broad diffusion peak at zero frequency \cite{mountain,rbspec}

\subsection{Short wavelength stability and causality}

Different aspects of the problem are probed in the short wavelength limit.
Letting $k\to \infty$ we see~\footnote{In order to be precise, we should state more clearly in what sense $k$ is large. However, the validity of the model, i.e., the physical scale(s) that the wavelength should be compared to are quite easy to work out from the final results, should one want to do so. Hence, we will not state the detailed condition here. } that \eqref{quartz} reduces to a quadratic for $\sigma^2$. 
This allows us to write down the solutions in closed form. Hence, it is relatively straightforward
to establish the conditions required for the stability of the system in this limit.
For infinitesimal wavelengths we have
\beq
A \sigma^4 - B \sigma^2 + C = 0 \ , \longrightarrow \ \sigma_\pm^2 = {1 \over 2A} \left[ B \pm \left(B^2 - 4AC \right)^{1/2} \right] \ .
\label{kvadrat}
\eeq
where 
\beq
A = m_*\tau> 0  \ ,
\eeq
in order for transverse perturbations to be stable. We also have
\beq
B = (p+\rho)\left( {\kappa\over n c_v} + c_s^2 \tau \right)-2\kappa T \alpha_s \ ,
\eeq
and
\beq
C =\kappa \left[ {p+\rho\over n} {c_s^2 \over c_{v}}  - T\alpha_s^2\right]  = \kappa \left( {p +\rho \over n}\right)  {c_s^2 \over c_p}  > 0\ .
\eeq

In order to guarantee stability for longitudinal perturbations, we need $\sigma^2$ to be real and positive. Given the quadratic formula and the fact that $A>0$ this implies that we must have $B^2 - 4AC > 0$. After some algebra, this leads to
\beq
\left( c_s^2 \tau - {\kappa \over n c_v} -{2\kappa T \alpha_s \over p+\rho} \right)^2 + {4\kappa T \alpha_s^2 \over p+\rho} \left( \tau - {\kappa T \over p+\rho} \right) + {4\kappa^2 T  \over (p+\rho) n c_v} \left(c_s^2-2\alpha_s\right)> 0 \ .
\label{discriminate}\eeq
The first two terms are positive, as long as  \eqref{tcon} is satisfied. Hence, the condition is guaranteed to be satisfied as long as $c_s^2>\alpha_s$.
As discussed later, this is certainly the case for degenerate matter. In cases where this simple condition is not satisfied, \eqref{discriminate} provides a complicated constraint on the relaxation time.
Finally, we must also have $B> 0$, which can be expressed as;
\beq
\tau > {\kappa \over  c_s^2} \left[ {2T \over p+\rho} \alpha_s - { 1 \over n c_v} \right] \ .
\label{ast}\eeq
This condition is identical to that given in eq. (146) of \cite{hl1} (obtained in the limit where $\alpha_i\to0$ and $1/\beta_0$ and $1/\beta_2$ both also vanish, c.f. \cite{herr3,maartens})

Let us now consider finite wavelengths. Letting $\sigma = \sigma_\pm+ \sigma_1/k$, where $\sigma_\pm$ solve \eqref{kvadrat}, 
and linearising in $1/k$, we find that
\beq
\sigma_1 =  {i(p+\rho) \over 2}  \left( {\sigma_\pm^2 - c_s^2 \over 2A \sigma_\pm^2 - B}\right) \ .
\label{sig1}\eeq
Since all quantities in this expression are already constrained to be real,  we need $\mathrm{Im}\ \sigma_1 \ge 0$ (for real $k$)
in order for the system to be stable.. From \eqref{kvadrat} we see
that
\beq
2A \sigma_\pm^2 - B = \pm\left|B^2 - 4AC \right|^{1/2} \ .
\eeq
This then leads to the final condition 
\beq
\sigma_-^2 \le c_s^2 \le \sigma_+^2 \ .
\label{fincon}\eeq
It is worth noting that this result is consistent with the notion that ``mode-mergers'' signal the onset of instability \cite{2stream}.

As the waves in the system must remain causal, we should insist that $\sigma^2< 1$. To ensure that this is the case, we adapt the strategy used by Hiscock and Lindblom \cite{hl1}. As \eqref{kvadrat} is a quadratic for $\sigma^2$ we can ensure that the roots are confined to the interval $0<\sigma^2 < 1$
(noting first of all that the roots are real since \eqref{discriminate} is satisfied). Given the $B$ and $C$ are both positive, the roots must be such that 
$\sigma^2>0$. Meanwhile we can constrain the roots to $\sigma^2 <1$ by insisting that
\beq 
A-B+C > 0 \ , 
\eeq 
and
\beq
A-2B > 0 \ .
\label{two}\eeq
Combining these inequalities with the positive discriminant, we can show that $A> B/2> C$. The first of the two conditions can be written
\beq
(1-c_s^2) \left[ \tau - {\kappa \over n c_v} \right] > {\kappa T (1-\alpha_s)^2 \over p+\rho }> 0 \ .
\label{hash}\eeq
Now, when combined with causality the condition \eqref{fincon} requires that $c_s^2 \le \sigma_+^2 < 1$. In other words, we must have 
$c_s^2< 1$, which means that \eqref{hash} implies that
\beq
\tau > {\kappa \over n c_v} \ .
\label{om6}\eeq
Comparing to the results of Hiscock and Lindblom \cite{hl1}, we recognize \eqref{hash} as their $\Omega_3>0$ condition (it is also eq. (4) of Herrera and Martinez \cite{herr2}), while \eqref{om6} 
corresponds to $\Omega_6>0$.

Meanwhile, the condition \eqref{two} can be written
\beq
(2-c_s^2) \tau > {\kappa \over n c_v} + {2\kappa T \over p+\rho} (1-\alpha_s) \ , 
\eeq
corresponding to eq. (148) of Hiscock and Lindblom.
Finally, $A>C$ leads to
\beq
\tau > {\kappa T \over p+\rho} + {\kappa c_s^2 \over n c_p}  \ .
\eeq
This corresponds to eq. (3) in Herrera and Martinez \cite{herr2}, which derives from eq. (147) of Hiscock and Lindblom \cite{hl1}.

This completes the analysis of the stability and causality of the system. We have arrived at a set of conditions on the thermal relaxation
time (and  related them to results in the existing literature). As long as these conditions are satisfied, the solutions to the problem should be well behaved. 

\subsection{The emergence of second sound}

So far we have considered the conditions that must be satisfied by a thermodynamical model in order
to ensure the stability and causality of both transverse and longitudinal waves. To complete the analysis of the problem, we
will now consider the nature of the solutions. Since the phase velocity $\sigma$ is obtained from a quartic,
we know that the problem has two (wave) degrees of freedom. This accords with the experience from superfluid systems
and experimental evidence for heat propagating as waves in low temperature solids. One of the solutions should be associated with 
the usual ``acoustic'' sound while the second degree of freedom will lead to a ``second sound'' for heat. 
We want to demonstrate how these features emerge within our model. 

In order to explore these features, it is natural to consider the large relaxation time limit. Taking the relaxation time 
$\tau$ to be long, the solutions to \eqref{kvadrat} take the form (up to, and including, order $1/\tau$ terms)
\beq
\sigma_+^2 \approx  c_s^2\left[ 1 +  {\kappa T \over (p+\rho) \tau} \left( 1 + {\alpha_s^2 \over c_s^4} \right)\right]  \ ,
\eeq
which could be rewritten using \eqref{cvrel}, and
and
\beq
\sigma_-^2 \approx  {\kappa \over n \tau c_p} \ .
\eeq
The first of these solutions clearly represents the usual sound, while the other solution provides the second sound. In the latter case, the 
deduced speed is exactly what one would expect \cite{joubook}.
It is easy to see that the first root will satisfy \eqref{fincon}, and the associated roots will  be unstable in the long relaxation time limit. 
Moreover, the second solution leads to stable roots provided
\beq
\tau \ge { \kappa \over n c_p c_s^2} \ .
\eeq
Basically, the finite wavelength condition implies that the second sound must propagate slower than the first sound. This is, indeed, what is measured in 
physical systems (like superfluid Helium). Moreover, it is easy to see that this condition must be satisfied in order for the long relaxation time approximation to be valid.

\begin{figure}[h]
\centering
\includegraphics[height=8cm,clip]{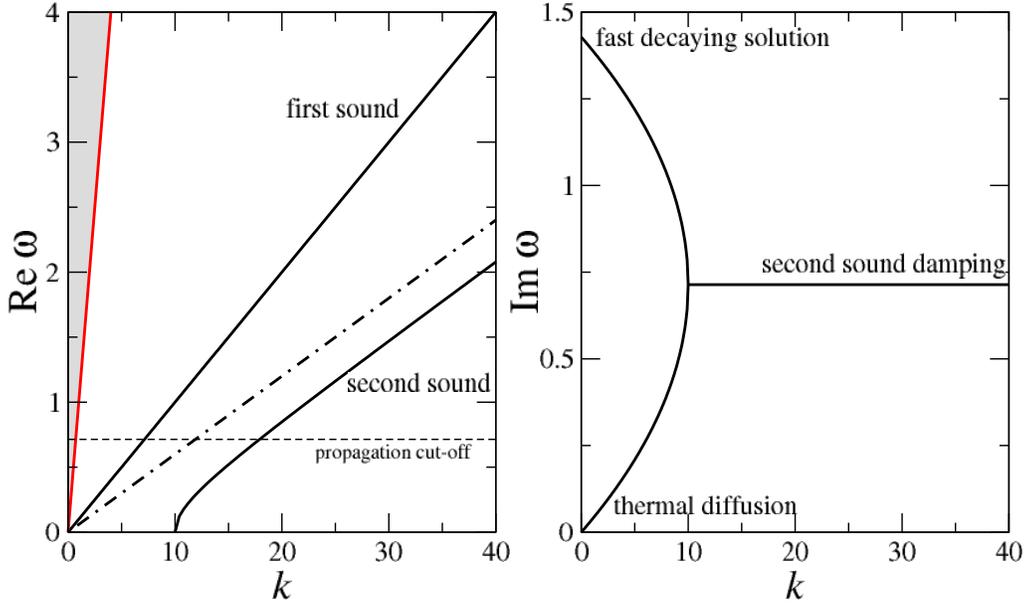}
\caption{This figure provides an illustration of the qualitative nature of the behaviour of heat conducting degenerate matter (as discussed in the main text), based on the consistent first-order relativistic model. The parameters have been chosen in such a way that the speed of sound is 10\% of the speed of light, while the second sound (at short wavelengths, large $k$) propagate at $1/\sqrt{3}$ of this.  The phase velocity of the waves is $\sigma=\mathrm{Re}\ \omega/k$ (left panel).The thermal relaxation time $\tau$ has been chosen such that the critical wavenumber at which the second sound emerges is $k=10$. At lengthscales larger than this, the corresponding roots are diffusive (have purely imaginary frequency), and in the very long wavelength limit ($k\to0$) we retain the expected thermal diffusion. The damping time follows from $1/\mathrm{Im}\ \omega$ (right panel).We also indicate the acausal region (grey are). 
The illustrated example is clearly both stable and causal. }
\label{heatplot}
\end{figure}

As a useful illustration of the properties of the model, let us consider the particular case of degenerate matter. In this case, which relates to 
electrons in both metals and white dwarfs (and also neutrons and protons in neutron stars), the two specific heats are almost identical;  
\beq
{c_v \over c_p} \approx 1 + O\left({k_B T \over \epsilon_F}\right)^2 \ ,
\eeq
where $k_B$ is Boltzmann's constant and $ \epsilon_F$ is the Fermi energi \cite{ashcroft}. This means that, for temperatures significantly below
the Fermi temperature, we can accurately assume that $\alpha_s \approx 0$. If we also assume that $\tau \gg \kappa T/(p+\rho)$ then the dispersion relation factorises and we have
\beq
\left( \sigma^2 - c_s^2 \right) \left( \tau \sigma^2 - {i \sigma \over k} - {\kappa \over n c_v} \right) = 0 \ . 
\eeq
That is, the four roots are
\beq
\sigma = \pm c_s \ , 
\eeq
and
\beq
\sigma = { i \over 2k\tau} \left[ 1 \pm \left( 1 - {4 \kappa \tau \over n c_v} k \right)^{1/2} \right] \ .
\label{2root}\eeq
The character of these roots is illustrated in Figure~\ref{heatplot}. We see that the ordinary sound exists at all wavelengths. Meanwhile, at short long wavelengths (small $k$) the remaining two roots are exponentially damped, i.e. diffusive in character. One root has a relatively slow decay, corresponding to the expected thermal diffusion, while the other root decays so rapidly that it is unlikely to be observable by experiment. Below a critical lengthscale 
(corresponding to $k=10$ in Figure~\ref{heatplot}) the second sound emerges as a result of the finite thermal relaxation time $\tau$. For very short lengthscales, heat signals will propagate as waves. However, as is clear from \eqref{2root} these solutions are always damped. In order to `propagate' the real part of the wave frequency must exceed the imaginary part (so that several cycles are executed before the motion is damped out). This boils down to the second sound propagating only for
\beq
k \gg \left({ \kappa \tau \over n c_v}\right)^{-1/2} \ .
\eeq
This result is interesting if we consider systems that become superfluid (see \cite{kaca} for an interesting discussion of models for relativistic superfluids). Suppose we consider a system which starts out in the diffusive regime (e.g. Helium above the superfluid transition temperature). When the system is cooled down through the relevant transition temperature, the (non-momentum conserving) particle collisions that give
rise to $\kappa$ are suppressed. In effect, the critical value of $k$ decreases and the system may enter the regime where the second sound can propagate
at macroscopic scales. The change to the basic nature of heat propagation is easily understood. 

At the end of the day, this relatively simple model demonstrates how the behaviour of a given physical system depends on the balance between the characteristic timescales. Obviously, we also need to keep in mind that real systems impose restrictions both for large $k$, as the fluid model 
breaks down when we approach the interparticle distance scale, and small $k$, when the wavelength becomes larger than the size of the system.

\section{Final remarks}

We set out to derive a consistent first-order model for relativistic heat conduction, in such a way that the theory remains cognizant of
its higher order origins. As we have demonstrated, this leads to a model that retains the thermal relaxation that is necessary if we want the problem to remain causal. What have we learned from this exercise? First of all, we have illustrated the problems associated with the original first-order models, e.g. that of Eckart \cite{eckart}. The conclusions are, obviously, not new but the discussion should lay to rest any suggestions that the coupling of the four-acceleration to the heat-flux is (somehow) problematic \cite{heatcoup1,heatcoup2,heatcoup3}. By considering the waves present in the new model, we have established that the system is both stable and causal provided
that some seemingly natural conditions are satisfied. This conclusion accords with the classic work by Hiscock and Lindblom \cite{hl1,hl2} (see also Olson and Hiscock \cite{olson}).
In fact,  
the conditions we have arrived at reproduce their key results. However, our analysis adds to previous work by discussing the emergence of the second sound and the nature of the associated solutions. This is a key point, especially if we are interested is relativistic superfluid systems. The analysis also demonstrates the intricate nature of these problems. It is easy to see how a model that may fail one, or several, of the derived conditions in some regime may nevertheless be valid for a different range of 
parameters. Hence, one really should consider the applicability of the chosen theory on a case-by-case basis. This is 
probably no more than should be expected from a phenomenological model.

Our results represent useful progress in this problem area, but one could obviously develop the theory further. A natural step would be to consider the various constraints that we have derived for detailed equations of state, e.g., matter coupled to phonons. It would also  be interesting to  consider applications of the first-order construction. While the model is restricted in the sense that it does not 
account for non-adiabatic effects, there is an exciting range of possible applications in astrophysics, cosmology and high-energy physics. Particularly interesting questions concern to what extent second sound effects are relevant in relativistic systems and the difference between first-order results and the, considerably more complex, 
second-order theories. 

\section*{Acknowledgements}

We would like to thanks Greg Comer and Lars Samuelsson for stimulating discussions.
NA acknowledges support from STFC via grant number ST/H002359/1. CSLM gratefully acknowledges support from CONACyT, 
and thanks QMUL for generous hospitality.

\bibliographystyle{aipproc}   

\end{document}